\documentstyle[preprint,prl,aps,epsf]{revtex}
\begin{document}
\tightenlines 
\narrowtext
\title{Gradient Symplectic Algorithms for 
Solving the Schr\"odinger Equation with Time-Dependent Potentials}
\author{ Siu A. Chin and C. R. Chen}
\address{Center for Theoretical Physics, Department of Physics,\\ 
Texas A\&M University, 
College Station, TX 77843}
\maketitle
\begin{abstract}

We show that the method of factorizing the evolution operator
to fourth order with purely positive coefficients,
in conjunction with Suzuki's method of implementing time-ordering of operators,
produces a new class of powerful algorithms for solving the Schr\"odinger 
equation with time-dependent potentials. When applied to the Walker-Preston 
model of a diatomic molecule in a strong laser field, these algorithms can 
have fourth order error coefficients that are three orders of magnitude smaller 
than the Forest-Ruth algorithm using the same number of Fast Fourier Transforms. 
When compared to the second order split-operator method, some of these algorithms 
can achieve comparable convergent accuracy at step sizes 50 times 
as large. Morever, we show that these algorithms belong to a one-parameter 
family of algorithms, and that the parameter  
can be further optimized for specific applications.

\end{abstract}

\section {Introduction}

Recently, we derived a new class of fourth order algorithms for solving both
classical\cite{chin97,chindon} and quantum dynamical\cite{chinchen} problems. 
These algorithms are based on factorizing the evolution operator 
${\rm e}^{\epsilon(T+V)}$ to fourth order with purely positive coefficients 
and require knowing the gradient of the force in the classical case and the 
gradient of the potential in the quantum case. The resulting algorithms are 
symplectic or unitary, respectively. While positive coefficients are 
absolutely necessary for simulating the diffusion process in Monte Carlo 
algorithms\cite{lang4,dmc4}, or doing imaginary time projections\cite{auer}, 
they are not essential in quantum or classical algorithms. 
Nevertheless, we have shown that this class of {\it gradient symplectic 
algorithms} is far superior to existing fourth order algorithms with 
negative coefficients\cite{chin97,chindon,chinchen}. In this work, using Suzuki's 
method\cite{texp} of implementing operator time-ordering, we prodouce a class 
of even more effective algorithms for solving quantum dynamical problems with
explicit time-dependent potentials. Despite the vast literature on this subject
\cite{feit,glasner,shen,takah,perskin,perskin2,grayver,zhucls,zhutdep,serna,blanes}, 
we believe our work has initiated a new direction in algorithm development. 
In the past, one labors assiduously to avoid higher order commutators. 
Here, we show that their inclusion can yield algorithms of great efficiency. 
Our algorithm 4A, to be discuss below, is the fastest fourth order 
algorithm known, needing only four Fast Fourier Transforms (FFT) 
per iteration. This is only twice the computational effort of the second 
order split-operator method\cite{feit,grayver,blanes}, but 
the algorithm can converge 
at time step sizes an order of magnitude larger. Our optimized
algorithms, when compared on an equal effort basis, have fourth order error 
coefficients that are three orders of magnitude smaller than 
Forest-Ruth's algorithm\cite{forest} and a factor of 30 smaller than McLachlan's 
algorithm\cite{mclach}; both are  
fourth order algorithms with negative coefficients.
																	
While these gradient symplectic algorithms are very efficient when the 
potential gradient is known analytically, they remain equally effective 
when the gradient is obtained numerically\cite{auer}. 
These algorithms are of particular interest in solving the time-dependent 
Schr\"odinger in a large 3D mesh. In 3D, even for a modest 
grid size of (256)$^3$, the number of mesh points already exceed 16 millions. 
If the wave function array is double precision and complex, its storage 
alone would have required 268MB. For such a large number of grid points, 
any vector-matrix multiplication would be prohibitively expensive and 
must be avoided. For our algorithms, the costliest computational step 
is just the use of FFT. The unitary character and the large time 
step acceptance of these algorithms make them ideal for doing long time 
quantum simulations.

The key problem in solving the Schr\"odinger equation with time-dependent 
potentials is the time ordering of operators. This problem is solved variously
in the literature by transforming it into a classical problem\cite{grayver,zhucls}, 
treating time as another ``spatial coordinate"\cite{perskin,perskin2,zhutdep}, 
introducing auxiliary variables\cite{serna}, etc.. Most end up with some 
sort of time derivative operator, but none has the simplicity 
of Suzuki's method\cite{texp} of directly implementing time-ordering via a 
{\it forward} time derivative operator. No time integration is necessary.
Since this work is less accessible, but is of special relevance to 
the operator factorization approach of deriving algorithms, we summarize 
it in some detail in the next section. In Section III, we apply Suzuki's
method and derive four gradient symplectic algorithms for solving the 
time-dependent Schr\"odinger equation. In Section IV, we use these 
algorithms to solve the Walker-Preston model\cite{walker} and compare 
their convergent properties with existing algorithms. In Section V, 
we derive one-parameter families of these algorithms and show that 
they can be further optimized for specific 
applications. Section VI summarizes our conclusions.

\section {Operator Decomposition of Ordered Exponentials}
 
For $H(t)$ a time-dependent operator, the evolution equation  
\begin{equation}
{\partial\over{\partial t}}\psi(t)=H(t)\psi(t),
\label{teq}
\end{equation}
has the operator solution
\begin{equation}
\psi(t+\Delta t)=T\Bigl(\exp\int_t^{t+\Delta t}H(s)ds\Bigr)\psi(t).
\label{expth}
\end{equation}
The time-ordered exponential not only has the conventional expansion
\begin{equation}
T\Bigl(\exp\int_t^{t+\Delta t}H(s)ds\Bigr)=1+\int_t^{t+\Delta t}H(s_1)ds_1
+\int_t^{t+\Delta t}ds_1\int_t^{s_1}ds_2H(s_1)H(s_2)+\cdots,
\label{texpexp}
\end{equation}
but also the more intuitive interpretation
\begin{eqnarray}
T\Bigl(\exp\int_t^{t+\Delta t}H(s)ds\Bigr)
=&&\lim_{n\rightarrow\infty}
T\Bigl({\rm e}^{
{{\Delta t}\over n}
\sum_{i=1}^{n}H(t+i{{\Delta t}\over n})}\Bigr),\nonumber\\
=&&\lim_{n\rightarrow\infty}
{\rm e}^{ {{\Delta t}\over n}H(t+\Delta t)}
\cdots
{\rm e}^{ {{\Delta t}\over n}H(t+{{2\Delta t}\over n})}
{\rm e}^{ {{\Delta t}\over n}H(t+{{\Delta t}\over n})}.
\label{torder}
\end{eqnarray}
There are many ways of solving the time-ordering problem. 
For this work on operator factorization,
we prefer Suzuki's method\cite{texp}, which directly implements
time ordering without any additional formalism\cite{perskin} or 
auxiliary variables\cite{serna}. 
 
Let $D$ denotes 
the {\it forward time derivative} operator
\begin{equation}
D={{\buildrel \leftarrow\over\partial}\over{\partial t}}
\label{ftsh}
\end{equation}
such that for any two time-dependent functions $F(t)$ and $G(t)$,
\begin{equation}
F(t){\rm e}^{\Delta t D}G(t)=F(t+\Delta t)G(t).
\label{fg}
\end{equation}
Suzuki's proved\cite{texp} that 
\begin{equation}
T\Bigl(\exp\int_t^{t+\Delta t}H(s)ds\Bigr)=\exp[\Delta t(H(t)+D)].
\label{tdecom}
\end{equation}
Using the more intuitive definition of time-ordering (\ref{torder}),
and invoking Trotter's formula,
one proof only requires two lines:
\begin{eqnarray}
\exp[\Delta t(H(t)+D)]
=&&\lim_{n\rightarrow\infty}\Bigr( {\rm e}^{ {{\Delta t}\over n}H(t)}
     {\rm e}^{ {{\Delta t}\over n}D}\Bigr)^n,\nonumber\\
=&&\lim_{n\rightarrow\infty}
{\rm e}^{ {{\Delta t}\over n}H(t+\Delta t)}
\cdots
{\rm e}^{ {{\Delta t}\over n}H(t+{{2\Delta t}\over n})}
{\rm e}^{ {{\Delta t}\over n}H(t+{{\Delta t}\over n})},
\label{altas}
\end{eqnarray}
where property (\ref{fg}) has been applied repeatedly and accumulatively.

For the widely applicable case of $H(t)=T+V(t)$, where only one of
the operator is explicitly dependent on time, the short time
evolution of (\ref{expth}) can be written using (\ref{tdecom}) as
\begin{equation}
\psi(t+\Delta t)
={\rm e}^{\Delta t[\widetilde T +V(t)]}\psi(t),
\label{tindd}
\end{equation}
which is just like the time-independent case but with
an effective $\widetilde T=T+D$. This suggests a two-step
approach of deriving time-dependent algorithms. First, decompose 
${\rm e}^{\Delta t[\widetilde T +V(t)]}$  in 
terms of ${\rm e}^{\Delta t \widetilde T}$ 
and ${\rm e}^{\Delta t V(t)}$ using any factorization scheme applicable
in the time-independent case. Next, since $[T,D]=0$, factorize exactly  
\begin{equation} 
{\rm e}^{\Delta t \widetilde T }
={\rm e}^{\Delta t D }{\rm e}^{\Delta t T }, 
\end{equation}
and incorporate all time-dependent requirements 
by applying (\ref{fg}). For example, a second order factorization of 
(\ref{tindd}) gives,
\begin{eqnarray}
{\cal T}^{(2)}_{A}
&&=
{\rm e}^{{1\over 2}\Delta t\widetilde T }
{\rm e}^{\Delta t V(t)}
{\rm e}^{{1\over 2}\Delta t\widetilde T },\nonumber\\
&&=
{\rm e}^{{1\over 2}\Delta t D }
{\rm e}^{{1\over 2}\Delta t T }
{\rm e}^{\Delta t V(t)}
{\rm e}^{{1\over 2}\Delta t D }
{\rm e}^{{1\over 2}\Delta t T },\nonumber\\
&&=
{\rm e}^{{1\over 2}\Delta t T }
{\rm e}^{\Delta t V(t+\Delta t/2)}
{\rm e}^{{1\over 2}\Delta t T },
\label{midpoint}
\end{eqnarray}
which is the well known midpoint algorithm for time-dependent problems. 
The other second order factorization gives the alternative second
order algorithm, 
\begin{eqnarray}
{\cal T}^{(2)}_{B}
&&=
{\rm e}^{{1\over 2}\Delta t V(t)}
{\rm e}^{\Delta t\widetilde T }
{\rm e}^{{1\over 2}\Delta t V(t)},\nonumber\\
&&=
{\rm e}^{{1\over 2}\Delta t V(t)}
{\rm e}^{\Delta t D }
{\rm e}^{\Delta t T }
{\rm e}^{{1\over 2}\Delta t V(t)},\nonumber\\
&&=
{\rm e}^{{1\over 2}\Delta t V(t+\Delta t)}
{\rm e}^{\Delta t T }
{\rm e}^{{1\over 2}\Delta t V(t)}.
\label{vv}
\end{eqnarray}
Thus, for $H(t)=T+V(t)$,
{\it the effect of time-ordering is to increment the time-dependence of each 
potential operator $V(t)$ by the sum of time steps of all the $T$ operators 
to its right}.

For the Schr\"odinger equation with a time-dependent potential,
the wave function is evolved forward in a short time $\Delta t$ by 
\begin{equation}
\psi(\epsilon)
={\rm e}^{\epsilon[\widetilde T +V(t)]}\psi(0),
\label{schexp}
\end{equation}
where $\epsilon=-i\Delta t$, $\widetilde T=\widetilde D +T$,
$\widetilde D=iD$, 
\begin{equation}
T=-{1\over{2\mu}}\sum_i{{\partial^2}\over{\partial x_i^2}}$\qquad{\rm and}\qquad 
$V(t)=V(x_i,t).
\end{equation} 
We will work in atomic units such that the kinetic energy 
operator has this standard form. Moreover, to do away with messy notations
involving $-i$, we will use $\epsilon$ as the time step variable everywhere. When
$\epsilon$ appears as the argument of the wave function or potential, it is to be 
understood that it denotes only the real time step variable $\Delta t$ 
without the factor $-i$. (In this way, algorithms can be directly applied to the 
classical case with $\epsilon$ purely real without change in form.) 
For conciseness, we will always
regard the present time step as time zero. Thus, the two second order algorithms
for solving the Schr\"odinger equation with step size $\Delta t$ can be denoted
simply as
\begin{eqnarray}
{\cal T}^{(2)}_{A}(\epsilon)
&&=
{\rm e}^{{1\over 2}\epsilon T }
{\rm e}^{\epsilon V(\epsilon/2)}
{\rm e}^{{1\over 2}\epsilon T },
\label{al2a}\\
{\cal T}^{(2)}_{B}(\epsilon)
&&=
{\rm e}^{{1\over 2}\epsilon V(\epsilon)}
{\rm e}^{\epsilon T }
{\rm e}^{{1\over 2}\epsilon V(0)}.
\label{al2b}
\end{eqnarray}

Since the kinetic energy operator is diagonal in 
momentum space, the operator ${\rm e}^{\epsilon T }$ can be implemented
as a vector-vector multiplication in Fourier space. Every occurrence of 
${\rm e}^{\epsilon T}$ requires two FFTs, one direct to Fourier space 
for the kinetic energy multiplication, and one inverse back to real space for 
the potential energy multiplication. To minimize the call for FFTs, one
favors algorithms with the fewest occurrence of the kinetic energy operator.  
  
\section {Gradient Symplectic Fourth Order Algorithms}

Following our two-step approach, we can transcribe
any time-independent factorization algorithm into a time-dependent algorithm. 
For example, the well known Forest-Ruth (FR) 
algorithm\cite{forest}(also discovered independently by Campostrini 
and Rossi\cite{camp},and Candy and Rozmus\cite{candy}) can now be transcribed to
solve time-dependent problems as
\begin{equation}
{\cal T}^{(4)}_{FR}(\epsilon)\equiv
{\rm e}^{v_3\epsilon V(\epsilon)}
{\rm e}^{t_3\epsilon T}
{\rm e}^{v_2\epsilon V(a_2\epsilon)}
{\rm e}^{t_2\epsilon T}
{\rm e}^{v_1\epsilon V(a_1\epsilon)}
{\rm e}^{t_1\epsilon T}
{\rm e}^{v_0\epsilon V(0)},
\label{alfr}
\end{equation} 
where, $s=2\,^{1/3}$,
\begin{equation}
v_0=v_3={1\over 2}{1\over{2-s}},\quad 
v_1=v_2=-{1\over 2}{{s-1}\over{2-s}},\quad
t_1=t_3={1\over{2-s}},\quad 
t_2=-{s\over{2-s}},
\label{vcof}
\end{equation}
and $a_1=t_1$, $a_2=t_1+t_2$. 
For easy identification, we adopt the convention of labelling 
the time step coefficients of operators $T$ and $V$ by $t_i$ and $v_i$ 
respectively, and denote the intermediate time {\it arguments} of $V$ by 
coefficients $a_i$. The coefficient of the first operator on the right 
will be denote by $v_0$ or $t_0$, followed by paired coefficients of
$v_i$ and $t_i$ for $i=1,2, \dots n$.
If one were to decompose ${\rm e}^{\epsilon (T+V)}$ only in terms of 
${\rm e}^{\epsilon T }$ and ${\rm e}^{\epsilon V }$, then
Forest-Ruth is the only fourth order algorithm
possible with 6 FFTs per iteration. 
The alternative algorithm with $T$ and $V$ interchanged, with appropriate
modifications of the $a_i$ coefficients, is also possible, but would
have required 8 FFTs. Notice that some of the coefficients are
negative, requiring backward propagation and evaluating the potential at
a time prior to the present. This is a consequence of Suzuki's ``no-go" 
theorem\cite{nogo}, which proved that beyond second order, 
${\rm e}^{\epsilon (T+V) }$ cannot be decomposed into a finite products 
of ${\rm e}^{t_i\epsilon T }$ and ${\rm e}^{v_i\epsilon V }$ with purely 
positive coefficients $t_i$ and $v_i$. Thus without exception,
all higher order factorization algorithms heretofore proposed in the 
literature\cite{glasner,shen,takah,zhucls,zhutdep,serna,blanes} 
contain negative coefficients. 

It is the search for positive coefficient 
factorizations schemes\cite{supos} that led one of us to derive
fourth order symplectic algorithms for solving classical\cite{chin97} 
and subsequently quantum dynamical\cite{chinchen} problems. 
To circumvent Suzuki's ``no-go" theorem, these factorization
schemes employ an additional operator,
\begin{equation} 
[V,[T,V]]={1\over\mu}\sum_i\biggl({{\partial V}\over{\partial x_i}}\biggr)^2,
\end{equation}
which is just the square of the gradient of
the potential\cite{supos}. This can be computed analytically or
numerically. For brevity, we will transcribe four previously derived
gradient algorithms\cite{chinchen} to their time-dependent form. 
In the next section, we will describe in more detail the one-parameter family of 
these algorithms. 

Our algorithm 4A\cite{chinchen} (See also Ref.\cite{suzukiab}), 
when applied to the time-dependent case of (\ref{schexp}), gives
\begin{equation}
{\cal T}_{A}^{(4)}(\epsilon)\equiv 
  {\rm e}^{ {1\over 6}\epsilon V(\epsilon)}
  {\rm e}^{ {1\over 2}\epsilon T} 
  {\rm e}^{ {2\over 3}\epsilon \widetilde V(\epsilon/2)} 
  {\rm e}^{ {1\over 2}\epsilon T} 
  {\rm e}^{ {1\over 6}\epsilon V(0)},
\label{foura}
\end{equation}
with $\widetilde V$ defined by
\begin{eqnarray}
\widetilde V(t)
&&=V(t)+{1\over 48}\epsilon^2[V(t),[\widetilde T,V(t)]],
\nonumber\\
&&=V(t)+{1\over 48}\epsilon^2[V(t),[T,V(t)]].
\label{superv}
\end{eqnarray}
Note that this is a crucial simplification, because $[V(t),[D,V(t)]]=0$\,!
Thus the addition of the forward time derivative operator $D$
causes no additional complication to the gradient term. 
( This is very fortunate, because if one were required to keep the alternative
double commutator $[\widetilde T,[V(t),\widetilde T]]$, this commutator 
would have given rise to complicated new terms involving 
$\partial V(t)/\partial t$ and $\partial^2 V(t)/\partial t^2\,$! )
With the introduction of the gradient term, algorithm 4A can
achieve fourth order acccuracy with only four FFTs. Note also that 
in order to minimize
the evaluation of the gradient term, the double commutator has
been placed at the center. This choice seems obvious, but there is
intrinsic freedom to redistribute the commutator term
among the three potential operator without affecting the fourth order
convergence of the algorithm. This is true for all algroithms described below.
We will assume that this redistribution can be done when necessary.     

Similarly algorithm 4B can be transcribed as 
\begin{equation}
{\cal T}_{B}^{(4)}(\epsilon)\equiv 
  {\rm e}^{t_2\epsilon T}
  {\rm e}^{\epsilon {1\over 2} \bar V(a_2\epsilon) } 
  {\rm e}^{t_1\epsilon   T}
  {\rm e}^{\epsilon {1\over 2} \bar V(a_1\epsilon) } 
  {\rm e}^{t_0\epsilon T},
\label{fourb}
\end{equation}
where
\begin{equation}
t_0=t_2={1\over 2}(1-{1\over{\sqrt 3}}),
\quad t_1={1\over{\sqrt 3}},
\quad a_1={1\over 2}(1-{1\over{\sqrt 3}}), 
\quad a_2={1\over 2}(1+{1\over{\sqrt 3}}),
\end{equation}
and with $\bar V$ given by
\begin{equation}
\bar V(t)=V(t)+{1\over 24}(2-\sqrt 3)\epsilon^2[V(t),[T,V(t)]]. 
\label{duperv}
\end{equation}
The time-dependent forms of algorithm 4C and 4D are respectively,
\begin{equation}
{\cal T}_{C}^{(4)}(\epsilon)\equiv 
  {\rm e}^{ {1\over 6}\epsilon T} 
  {\rm e}^{ {3\over 8}\epsilon V(5\epsilon/6)}
  {\rm e}^{ {1\over 3}\epsilon T} 
  {\rm e}^{ {1\over 4}\epsilon\widetilde V(\epsilon/2)}
  {\rm e}^{ {1\over 3}\epsilon T} 
  {\rm e}^{ {3\over 8}\epsilon V(\epsilon/6)}
  {\rm e}^{ {1\over 6}\epsilon T},
\label{chinc}
\end{equation}
\begin{equation}
{\cal T}_{D}^{(4)}(\epsilon)\equiv 
  {\rm e}^{{1\over 8}\epsilon \widetilde V(\epsilon)}
  {\rm e}^{{1\over 3}\epsilon T} 
  {\rm e}^{{3\over 8}\epsilon V(2\epsilon/3)}
  {\rm e}^{{1\over 3}\epsilon T}
  {\rm e}^{{3\over 8}\epsilon V(\epsilon/3)}
  {\rm e}^{{1\over 3}\epsilon T} 
  {\rm e}^{{1\over 8}\epsilon \widetilde V(0)},
\label{chind}
\end{equation}
where $\widetilde V(t)$ is as defined by (\ref{superv}).
The number of FFT required by each algorithm is given
in Table \ref{tabal}. 

\section {Solving the Walker-Preston model}

To demonstrate the effectiveness of these new algorithms, we use them to
solve the Walker and Preston model\cite{walker} of a diatomic molecule 
in a strong laser field. Since this problem has been used by many 
authors\cite{perskin2,grayver,serna,blanes} to test 
their time-dependent algorithms, it is an excellent choice for 
comparing our algorithms. The model is defined by the one dimensional
Hamiltonian (in atomic units),  
\begin{equation}
H=-{1\over{2\mu}}{{\partial^2}\over{\partial x^2}}+V(x,t),
\label{model}
\end{equation}
with 
\begin{equation}
V(x,t)=V_0(1-{\rm e}^{-\alpha x})^2+Ax\cos(\omega t),
\label{hamop}
\end{equation}
and where $\mu=1745$, $V_0=0.2251$, $\alpha=1.1741$, $A=0.011025$, and
$\omega=0.01787\,$. The wave function is initially chosen to be the
Morse oscillator ground state, which has the corresponding ground 
state energy $E_0=(\omega_0/2)[1-\omega_0/8V_0]$ with 
$\omega_0=\alpha\sqrt{2V_0/\mu}$. 
In conformity with the above authors, we discretize the wave function 
$\psi(x,t)$ using 64 uniform points at $x_k=-0.8+k\Delta x$, with 
$\Delta x=0.08\,$. The natural time scale defined by the laser oscillation
frequence is $\tau=2\pi/\omega=351.6\,$.

\begin{figure}
\noindent
\vglue 0.2truein
\hbox{
\vbox{\hsize=7truein
\epsfxsize=6truein
\leftline{\epsffile{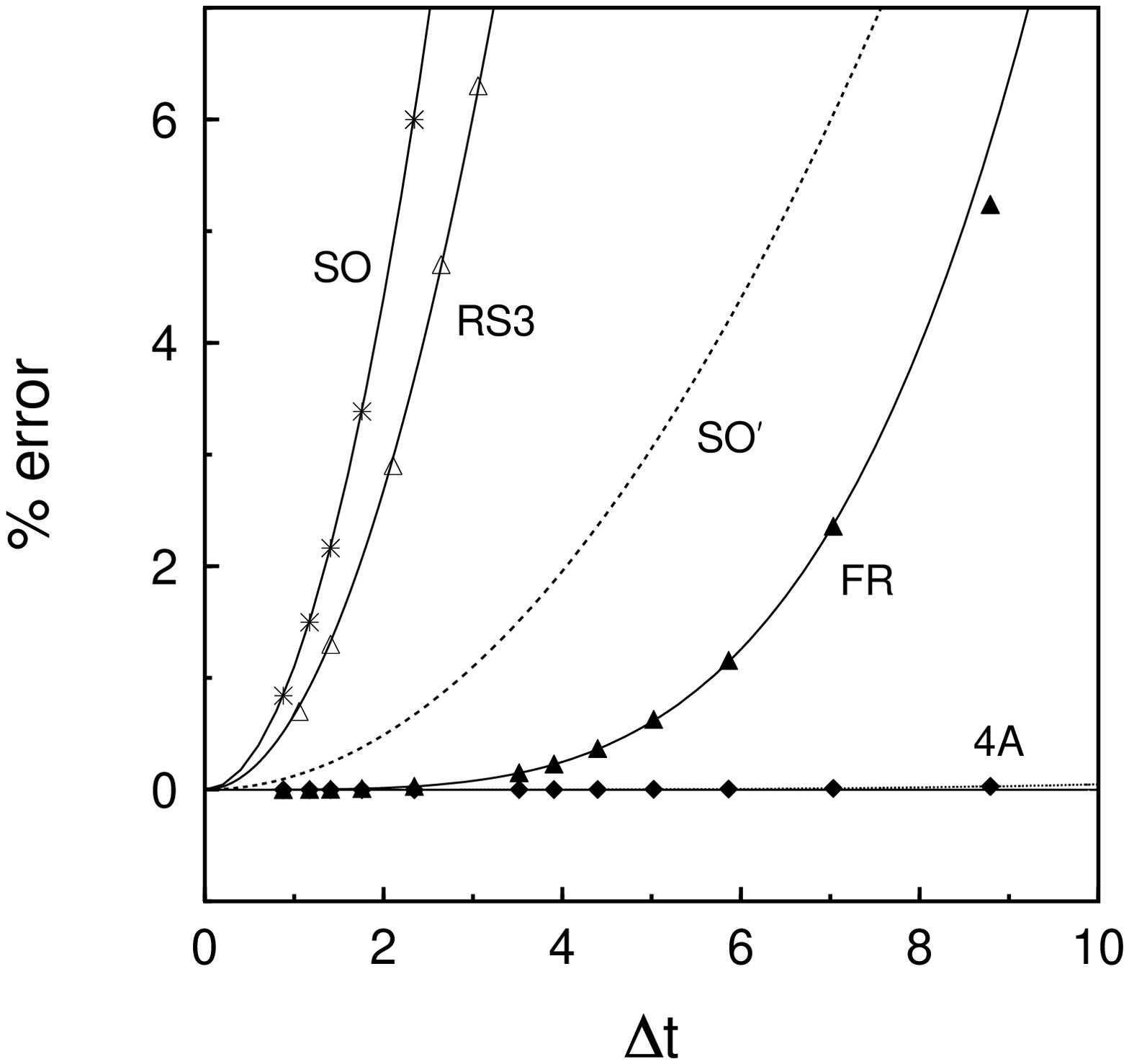}}
}
}
\vglue 0.3truein
\caption{The percentage energy error at $t=1000\tau$ for
various algorithms in solving the Walker-Preston model. 
The plotting symbols
are results of calculations. The lines are second or fourth
order fits. SO is the split-operator algorithm (\ref{al2b}), 
SR3 is one of Gray and Verosky's algorithm, 
FR is the Forest-Ruth algorithm (\ref{alfr}), and 4A is the
gradient algorithm (\ref{foura}). SO$^\prime$ is an equal computational 
effort version of SO for comparing with SR3 and FR.
See text for details.
}
\label{fone}
\end{figure}

In Fig.\ref{fone}, the percentage errors of the total energy, 
$100|E(1000\tau)-E_{con}|/E_{con}$, is plotted as a function 
of the step size $\Delta t$ used in the calculation. 
$E_{con}=5.029155E_0$ is the converged value at very small time steps. 
The plotting symbols indicate calculated results. The lines are
monomial fits in either $(\Delta t)^2$ or $(\Delta t)^4$.
SO is our calculation using the second order split-operator algorithm  
(\ref{al2b}). RS3 is Gray and Verosky's best convergent
results\cite{grayver}. They use a third order algorithm, but the error is 
degraded by the Magnus approximation to second order. 
Both can be well-fitted by a single quadratic $b_i(\Delta t)^2$, 
with coefficients $b_{SO}=1.1$ and $b_{RS3}=0.67\,$. These are plotted
as lines running through the data. The convergent range
for both is below $\Delta t<\tau/100=3.52$. Algorithm RS3 requires three times
as much effort as SO. For the same amount of computational effort, one can run
SO three times at $\Delta t/3$ and gain an reduction of
$(1/3)^2$ in its error coefficient.	This projected 
convergence curve $(b_{SO}/9)(\Delta t)^2$
is plotted as a broken line and labelled as SO$^\prime$. For the same effort
(6 FFTs) one can also run the Forest-Ruth algorithm and obtain results shown
as solid triangles, with a fitted line
$9.7\times 10^{-4}(\Delta t)^4$.
This equal effort comparison clearly demonstrates the greater efficiency
of fourth order algorithms. For errors in the range of 0.1 to 0.01 percent,
FR's time step size can be 4 to 6 times as large as SO$^\prime$'s.
This ratio would increase if greater accuracy is required.  
Algorithm 4A, which requires only 2/3 the effort of FR, can be fitted by
$4.8\times 10^{-6}(\Delta t)^4$. This fit is plotted as a dotted line 
barely visible above the zero error line over the range of the plot.

To discuss the convergence of our gradient symplectic algorithms, we
greatly expanded the plotting range in Fig.\ref{ftwo}. Here, we plot directly
the convergence of $E(1000\tau)/E_0$. The plotting symbols indicate 
calculated results. We retained SO and FR for comparison. The SO result 
is well-fitted by $E_{con}/E_0+0.054(\Delta t)^2$.
Algorithm 4C and 4D yielded indistinguishable results, despite the fact
that 4D uses only 6 FFTs, two less than 4C. Obviously one should not
use algorithm 4C in the present case; we only included it here for
completeness. Since the FR algorithm is known to have rather 
large errors, we have also implemented McLachlan's fourth order 
algorithm\cite{mclach} and obtained results labelled as M. This algorithm 
requires 8 FFTs per iteration and is the best algorithm tested 
by Sanz-Serna and Portillo\cite{serna}. Both FR and M are examples of 
fourth order algorithms with negative coefficients. 
The step size convergence of all fourth order algorithms can be 
very well-fitted by $E_{con}/E_0+d_i(\Delta t)^4$. These are plotted as lines 
going through data points. The coefficient $|d_i|$ for each algorithm is 
listed in Table \ref{tabal}. Since the computational effort per time
step is proportional to the number of FFTs, $N_i$, the error
per unit effort, taking into account the variation of step size with
effort, would be proportinal to $|d_i|N_i^4$, {\it i.e.}, this
is a measure of error of equal computational effort for all fourth order 
algorithms. Again, for example, algorithm 4A only requires half the number of
of FFTs as McLachlan's algorithm. At a given $\Delta t$ in running McLachlan's 
algorithm, one can execute algorithm 4A twice at $\Delta t/2$
and reduce its error by a factor of $(1/2)^4$. Thus despite the
appearance in Fig.\ref{ftwo}, algorithm 4A actually has a much smaller error per
unit effort than McLachlan's algorithm. We normalize this {\it equal effort} 
error to the value of FR and define the normalized,
equal-effort error coefficient as $\delta_{eq}=|d_i/d_{FR}|(N_i/N_{FR})^4$.
This value for each algorithm is listed in Table \ref{tabal}.
Thus, excluding 4C, our gradient algorithms are roughly
a factor of $10^3$ smaller in error than FR and a factor of 3 to 8 
smaller than McLachlan's algorithm. A even more useful measure is to consider
$\tau_{eff}=\delta_{eq}^{-1/4}$, which would give the time step size relative to RF 
for achieving the same error with equal effort. 
This is also given in Table \ref{tabal}. From examining Fig.\ref{fone} 
and \ref{ftwo}, it certainly seems reasonable that our algorithms can 
converge at time steps nearly 5 times as large as RF's and on the order of 
30 times as large as SO$^\prime$'s. We will discuss further optimized
algorithms 4ACB and 4BDA in the next section.     
  
\section {One-Parameter Family of Algorithms}

Algorithms 4A and 4C are special cases of the more
general algorithm 
\begin{equation}
{\cal T}_{ACB}^{(4)}(\epsilon)\equiv 
  {\rm e}^{ t_3\epsilon T} 
  {\rm e}^{ v_3\epsilon V(a_3\epsilon)}
  {\rm e}^{ t_2\epsilon T} 
  {\rm e}^{ v_2\epsilon\widetilde V(a_2\epsilon)}
  {\rm e}^{ t_1\epsilon T} 
  {\rm e}^{ v_1\epsilon V(a_1\epsilon)}
  {\rm e}^{ t_0\epsilon T}.
\label{algac}
\end{equation}
By use of a Mathematica program\cite{lang4,dmc4} to symbolically combine
exponentials of operators, it is easy to determine these
positive coefficients to yield a fourth	order algorithm. For the above
form of the algorithm, there is one free parameter, which we will take 
to be $t_0$. Given $t_0$, the rest of the coefficients are:
\begin{equation}
t_1=t_2={1\over 2}-t_0,\quad
t_3=t_0,\quad 
v_1=v_3={1\over 6}{1\over{(1-2 t_0)^2}},\quad 
v_2=1-(v_1+v_3),\quad
\label{cofac}
\end{equation}
where $\widetilde V$ here is given by
\begin{equation}
\widetilde V(t)
=V(t)+{u_0\over v_2}\epsilon^2[V(t),[T,V(t)]],
\label{vtac}
\end{equation}
with $a_1=t_0$, $a_2=1/2$, $a_3=1-t_0$, and 
\begin{equation}
u_0={1\over 12}\biggl[1-{1\over{1-2t_0}}+{1\over{6(1-2t_0)^3}}\biggr].
\label{acofac}
\end{equation}
For $t_0=0$, one has algorithm 4A. 
For $t_0=1/6\approx 0.167$, one recovers algorithm 4C. Thus one can change 
continuously from algorithm 4A to 4C, and beyond, by ``dialing" the 
parameter $0\leq t_0\leq{1\over 2}(1-{1\over{\sqrt 3}})\approx 0.21$. 
At the upper limit of $t_0= {1\over 2}(1-{1\over{\sqrt 3}})$,
$v_2=0$, and the algorithm becomes a variant of algoirthm 
4B, where the commutator term remained at the center rather than equally 
distributed between the two potential operators at positions $v_1$ and $v_3$.
We shall refer to this algorithm as 4B$^\prime$.
Algorithm 4B$^\prime$ can be continuously transformed to 4B by redistributing 
the commutator term from the center to both sides. For example, one can
multiply the central commutator term by a factor $1-\alpha$ and add 
$\alpha/2$ times the commutator term to each potential operator on the side.
As $\alpha$ ranges from 0 to 1, algorithm 4B$^\prime$
is continuously transformed to 4B. When $\alpha$ reaches 1, the central commutator 
disappears and the number of FFTs collapses from 8 to 6. Thus both algorithms 
4A and 4B are singular end points of this general algorithm with discontinuous
changes in the numbers of FFT.

Algorithm 4B and 4D are special cases of the general algorithm
\begin{equation}
{\cal T}_{BDA}^{(4)}(\epsilon)\equiv 
  {\rm e}^{ v_3\epsilon V(\epsilon)}
  {\rm e}^{ t_3\epsilon T} 
  {\rm e}^{ v_2\epsilon \widetilde V(a_2\epsilon)}
  {\rm e}^{ t_2\epsilon T} 
  {\rm e}^{ v_1\epsilon \widetilde V(a_1\epsilon)}
  {\rm e}^{ t_1\epsilon T}
  {\rm e}^{ v_0\epsilon V(0)},
\label{algbd}
\end{equation}
all requiring 6 FFTs. Here, we choose
$t_1$ as the free parameter. The coefficients are then
\begin{equation}
t_3=t_1,\quad
t_2=1-2t_1,\quad
v_3=v_0={{6t_1(t_1-1)+1}\over{12(t_1-1)t_1}},\quad 
v_2=v_1={1\over2}-v_0.
\label{cofbd}
\end{equation}
Here $\widetilde V$ is given by
\begin{equation}
\widetilde V(t)
=V(t)+{u_0\over v_2}\epsilon^2[V(t),[T,V(t)]],
\label{vtbd}
\end{equation}
with $a_1=t_1$, $a_2=1-t_1$, and
\begin{equation}
u_0={1\over 48}\biggl[{1\over{6t_1(1-t_1)^2}}-1\biggr].
\label{bdofac}
\end{equation}
Algorithm 4B corresponds to setting $v_0=0$ and selecting the smaller
of the two quadratic solution, 
$t_1={1\over 2}(1-{1\over{\sqrt 3}})$. When $t_1= 1/3$,
one obtains a variant of algorithm 4D, which we will denote as 
4D$^\prime$. Again, one can transform 4D$^\prime$ to 4D continuously by
distributing the commutator term at positions $v_1$ and $v_2$ 
to $v_0$ and $v_3$. However, in order to keep the number of gradient terms at 
a minimum, we will not bother with this refinement here. 
For positive coefficients, we must have 
${1\over 2}(1-{1\over{\sqrt 3}})\leq t_1\leq {1\over 2}$. 
At the upper limit of $t_1= {1\over 2}$, $t_2=0$, and the algorithm 
collapses back to algorithm 4A. Thus, there are two continuous families of
algorithms, the C-type requiring eight FFTs and the D-type 
requiring six. They are joined at both ends by
algorithms 4A and 4B with discontinuous changes in 
the numbers of FFTs. (Recently, Omelyan, Mryglod and Folk\cite{folk} have
considered the one-parameter factorization schemes from 4A to 4C
and 4B to 4D separately, without realizing that they can be further extended
and joined into one.)

\begin{figure}
\noindent
\vglue 0.2truein
\hbox{
\vbox{\hsize=7truein
\epsfxsize=6truein
\leftline{\epsffile{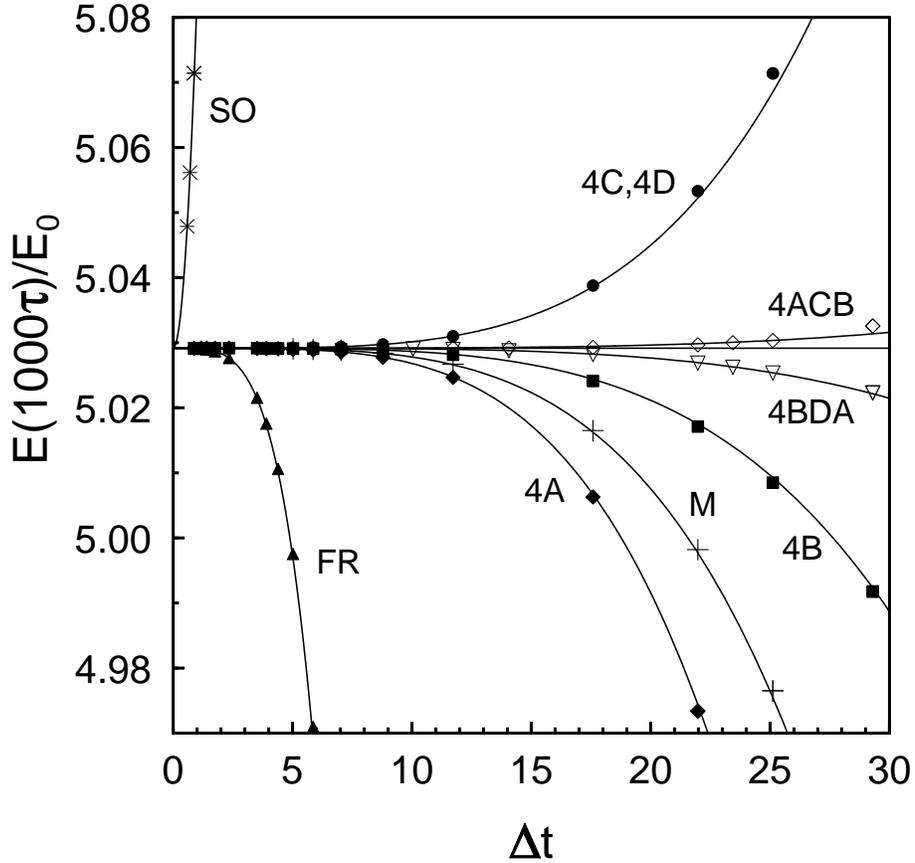}}
}
}
\vglue 0.3truein
\caption{
The long time energy convergence error for various algorithms 
in solving the Walker-Preston model. $E_0$ is the Morse
oscillator's ground state energy. M is McLachan's fourth
order algorithm. The gradient algorithms 4A, 4B, 4C, 4D are defined by
(\ref{foura}),(\ref{fourb}),(\ref{chinc}), and (\ref{chind}) respectively.
The optimized one-parameter algorithms 4ACB and 4BDA are described by
(\ref{algac}) and (\ref{algbd}). See text for details.
} 
\label{ftwo}
\end{figure}

Fig.\ref{ftwo} shows that the fourth order convergence error is negative 
for algorithm 4A and 4B, and positive for 4C and 4D. Since algorithm 4A can be 
continuously changed to 4C, and algorithm 4B be transformed to 4D, there 
must be parameter values such that this fourth order error can 
be made to vanish. This immediately suggests a simple strategy for 
further optimization: for each application, one can first minimized the 
convergence error for a {\it short} time run with respect to  
$t_0$ in ${\cal T}_{ACB}^{(4)}$ or $t_1$ in ${\cal T}_{BDA}^{(4)}$. One can 
then use the algorithm at that value and at a large time step, to do 
long time simulations. Since the error term is generally time dependent, 
there is no guarantee that when it is small for one time, it will remain 
small for all time. For the important case of periodic time-dependence, 
one can hope to reduce the bound within which the error can fluctuate. 
This seems to work for the present case.

\begin{figure}
\noindent
\vglue 0.2truein
\hbox{
\vbox{\hsize=7truein
\epsfxsize=6truein
\leftline{\epsffile{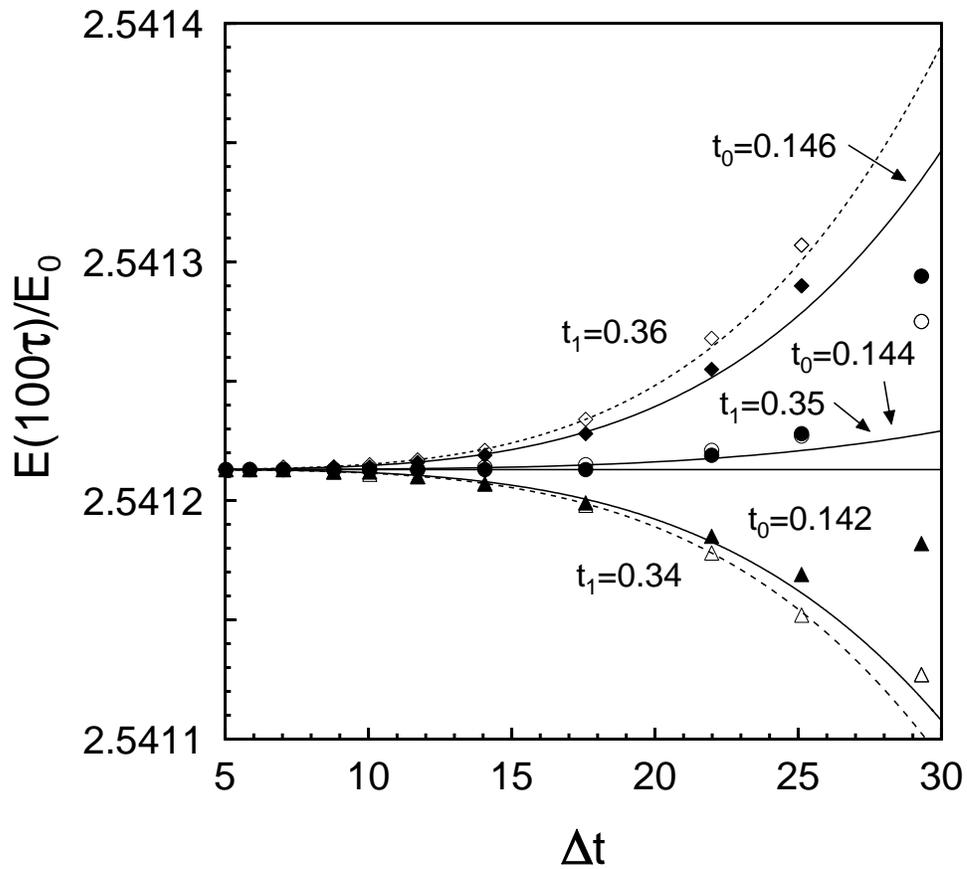}}
}
}
\vglue 0.3truein
\caption{The short time convergence error as a function of the
parameters $t_0$ and $t_1$. The solid symbols fitted by
solid lines are results
for algorithm 4ACB at three values of $t_0$. The hollow symbols
with broken lines are results for algorithm 4BDA at three values of $t_1$.
}
\label{fthree}
\end{figure}

From Fig. \ref{ftwo}, as the algorithm is changed continuously from 4A to 
4C and back to 4B, one should observe two zero error crossings. 
In varying $t_0$ in ${\cal T}_{ACB}^{(4)}$, we only see one. 
This suggests that the 
second crossing is when 4B$^\prime\rightarrow $4B, which we did not
consider in this work. The one zero crossing we observed is at $t_0=0.144$, 
precisely between 4A($t_0$=0) and 4C ($t_0$=1/6). 
A short run ($t=100\tau$) to determine the optimal value for $t_0$ 
is shown as solid symbols with solid fitting lines in Fig.\ref{fthree}. 
For $t_0=0.142$ and $t_0=0.146$, the convergence errors can be well fitted
by fourth order lines as shown. At the value of $t_0=0.144$, the
fourth order coefficient is an order of magnitude smaller. We deem it sufficient
to determine the zero-crossing parameter value to three-digit accuracy. 
The resulting long run is shown as 4ACB in Fig.\ref{ftwo}.
As shown in Table \ref{tabal}, $\delta_{eq}$ in this case
is reduce by an order of magnitude and $\tau_{eff}$ lengthened to 8.5$\,$.   
The convergence error curve for 4ACB remained very flat
in going from $t=100\tau$ to $t=1000\tau$.

In the case of ${\cal T}_{BDA}^{(4)}$, algorithm 4D$^\prime$ is sufficiently 
dissimilar to 4D such that the convergence error remained negative as 4B is 
morphed to 4D$^\prime$ at $t_1=1/3$. However, as $t_1$ increases above $1/3$, 
the convergence error moves up and turns positive. Since ${\cal T}_{BDA}^{(4)}$
ends up at 4A, the error must cross zero again on its way down. 
The first crossing is at $t_1=0.35\,$. This is shown in Fig.\ref{fthree} with 
hollow symbol connected by fitted, broken lines. The second crossing is near 
$t_1=0.46$, but we do not bother to show this similar result. 
The long-run results using $t_1=0.35$ in ${\cal T}_{BDA}^{(4)}$
is shown on Fig.\ref{ftwo} as 4BDA. As listed in Table \ref{tabal}, its equal 
effort error and effective time step factor are virtually identical to those 
of 4ACB. In both cases, without any additional computational effort, we 
have reduced the error by a factor of five and lengthen
the time step size by a factor 1.5$\,$.
Both algorithms can achieve 10$^{-4}$ accuracy in total energy
at a time step size 50 times as large as SO$^\prime$.  

A key advantage of these factorization algorithms is that one can
obtain analytically their energy error terms by use of Mathematica. 
For fourth order gradient algorithms, there are four non-vanishing 
error operators in the Hamiltonian,
\begin{equation}
\Delta E^{(4)}=\epsilon^4(e_1[VTVTV]+e_2[TTVTV]+e_3[VTTTV]+e_4[TTTTV]), 
\label{er}
\end{equation}
where $e_i$ are coefficient functions depending on $t_0$ or $t_1$,
and $[ABC]\equiv[A,[B,[C]]]$, etc.. The zero-crossing values of
$t_0$ in ${\cal T}_{ACB}^{(4)}$ and  $t_1$ in ${\cal T}_{BDA}^{(4)}$, 
are closely matched to the crossing points of the first
two error coefficients, $e_1=e_2$. For 4ACB and 4BDA, the predicted
zero-crossing values from solving $e_1-e_2=0$ numerically  
are $t_0=0.14233$ and $t_1=0.35023$ respectively, in excellent 
agreement with empirical values. It seems that out of the four error operators,
only two are dominant and they are opposite in sign. 
It should be noted that these crossing points have nothing to do with
the minimum of $\sum_{i=1}^4 e_i^2$. These algorithms are not optimized by
simply minimizing the sum of squares of error coefficients\cite{folk}. 
Further discussions on the above one-parameter family of gradient 
algorithms will be presented in a separate publication.   
 
\section {Conclusions}

In this work, we have derived a new class of fourth order algorithms
for solving the time-dependent Schr\"odinger equation with explicit
time-dependent potentials. This class of algorithms is characterized
by having only positive factorization coefficients and can achieve great
efficiency by knowing the gradient of the potential. For the Walker-Preston 
model, on an equal effort basis, the convergence errors of these algorithms 
can be 5000 and 30 times smaller than negative-coefficient algorithms such 
as the Forest-Ruth and the McLachlan algorithm, respectively. 
These gradient algorithms should be further tested on more realistic
problems where the gradient of potential may have to be computed numerically.

Our discovery of the one parameter family of gradient algorithms illustrates
the power of the operator factorization approach of solving any evolution
equation. The ability to tailor a specific algorithm for a particular application
reflects great versatility of our method. Our work encourages further systematic
study of these and other parametrized families of algorithm and their 
optimization. In particular, one should explore the freedom in 
redistributing the commutator term\cite{jang} in going from 4B to 
4B$^\prime$ and from 4D to 4D$^\prime$. We have not done so here in order
to concentrate on algorithms with the minimum number of gradient evaluations.
With the development of these powerful gradient algorithms for solving the 
Schr\"odinger equation directly, we can see no practical advantage in solving 
the corresponding classical problem\cite{grayver} using classical methods. 
This is in agreement with Sanz-Serna and Portillo's earlier assessment\cite{serna}. 

This work spurs further interest in finding six and higher order factorization 
algorithms with purely positive coefficients. Despite intense effort, we have
yet to find a sixth order factorization scheme with positive coefficients. 
It is most likely that one doesn't exist, even with the inclusion of the 
double commutator term. A recent work\cite{folk} reached the same conclusion, 
but offered no proof either. Such a proof, if exists, would make these 
fourth order gradient algorithms unique. There would be no higher order 
generalizations. This would raise many other interesting questions, such as 
what other operators are necessary for a higher order positive
time step factorization, what would be the optimal sixth order algorithm with
minimal of negative coefficients, etc.. All such investigations would deepen
our understanding of the operator factorization method of solving evolution
equations.

\acknowledgements
This work was supported, in part, by the National Science Foundation
grants No. PHY-0100839 to SAC.


\ifpreprintsty\newpage\fi
\ifpreprintsty\newpage\fi
\ifpreprintsty\newpage\fi
\begin{table}[t]
\caption[]{ 
Equal computational effort comparison of all fourth order
algorithms discussed in this work. All algorithms except
FR and M are our new time-dependent algorithms. $N_i$ is
the numbers of FFTs per execution for each
algorithm, $d_i$ is the corresponding fourth order
error coefficient, $\delta_{eq}$ is the equal-effort fourth order
error coefficient normalized to FR's value, and $\tau_{eff}$ is
the effective time-step size relative to FR's value, {\it e.g.}, 
for the same amount of effort, algorithm 4ACB can
achieve the same convergence error of FR at a time step 8.5 times
as large.  
}
\begin{center}
\begin{tabular}{|c|c|c|c|c|c|c|c|c|} 
              &FR    &M     &4A   &4B    &4C   &4D    &4ACB     &4BDA  \\ \hline
$N_i$           &6     &8     &4    &6     &8    &6     &8        &6     \\ \hline
$|d_i|$ &$5.0\times 10^{-5}$&$1.3\times 10^{-7}$&$2.4\times 10^{-7}$&$0.5\times 10^{-7}$&$1.0\times 10^{-7}$&$1.0\times 10^{-7}$&$3.0\times 10^{-9}$&$9.5\times 10^{-9}$\\ \hline
$\delta_{eq}$ &1     &$8.2\times 10^{-3}$&$0.95\times 10^{-3}$&$1.0\times 10^{-3}$&$6.3\times 10^{-3}$&$2.0\times 10^{-3}$&$1.9\times 10^{-4}$&$1.9\times 10^{-4}$\\ \hline
$\tau_{eff}$  &1     &3.3   &5.7  &5.6   &3.5  &4.7   &8.5      &8.5     \\ 
\end{tabular}
\end{center}
\label{tabal}
\end{table}
\end{document}